\documentclass[12pt]{article}
\usepackage{amsmath}
\usepackage{graphicx,psfrag,epsf}
\usepackage{enumerate}
\usepackage[square,sort,nonamebreak,comma]{natbib}
\pdfoutput=1
\usepackage{float}
\usepackage{bm}
\usepackage{dsfont}

\usepackage{algorithm}
\usepackage[noend]{algpseudocode}

\makeatletter
\def\BState{\State\hskip-\ALG@thistlm}
\makeatother

\usepackage{mathtools} 
\usepackage{scalerel}
\newcommand{\widesim}[2][1.5]{
  \mathrel{\overset{#2}{\scalebox{#1}[1]{$\sim$}}}}

\newcommand{\blind}{0}

\begin{document}

\def\spacingset#1{\renewcommand{\baselinestretch}%
{#1}\small\normalsize} \spacingset{1}


\if0\blind
{
  \title{\bf fMRI group analysis based on outputs from Matrix-Variate Dynamic Linear Models.}
  \author{Johnatan Cardona Jim\' enez\hspace{.2cm}\\
    Institute of Mathematics and Statistics, University of S\~ao Paulo, Brazil\\}
  \maketitle
} \fi

\if1\blind
{
  \bigskip
  \bigskip
  \bigskip
  \begin{center}
    {\LARGE\bf Title}
\end{center}
  \medskip
} \fi

\bigskip
\begin{abstract}

In this work, we describe in more detail how to perform fMRI group analysis using inputs from modeling fMRI signal using Matrix-Variate Dynamic Linear Models (MDLM) at the individual level. After computing a posterior distribution for the average group activation, the three algorithms (FEST, FSTS, and FFBS) proposed from the previous work by \cite{jimnez2019assessing} can be easily implemented. We also propose an additional algorithm, which we call AG-algorithm, to draw on-line trajectories of the state parameter and therefore assess voxel activation at the group level. The performance of our method is illustrated through one practical example using real fMRI data from a "voice-localizer" experiment.

\end{abstract}

\noindent%

\spacingset{1.45}
\section{Introduction}
\label{sec:intro}

Statistical fMRI group analysis is a procedure commonly used by practitioners in the neuroscience community. One of the aims of this type of analysis can be either identify patterns of brain activation for a specific group of individuals who suffer some types of illnesses (or conditions) or just for the general understanding of the functioning of the human brain under specific types of stimuli. One example of the former is the study performed by \cite{batistuzzo2015reduced}, where an fMRI group analysis is performed in order to characterized prefrontal activation under specific experimental conditions in pediatric patients who suffer from obsessive-compulsive disorders. On the other hand, one good example of studies which explores mechanisms of human brain activation is the work developed by \cite{pernet2015human}, where an fMRI group analysis is performed in order to characterize specific areas of the temporal cortex involved in the perception of the human voice. Those are just two examples, from thousands of studies that have been published since the arrival of the MRI scanners.
To carry on this type of fMRI group analysis there are some packages available, such as \textbf{FSL} \citep{jenkinson2012fsl} and \textbf{SPM} \citep{penny2011statistical}, with implementations of different types of statistical analysis. The most common statistical technique reported in the majority of fMRI studies is the so-called General Linear Model (GLM), which is noting more than a normal linear model.  The GLM is used only at the individual stage, where it is fitted on every voxel (voxel-wise approach) for every subject in the sample, and its estimated size effects (and sometimes its standard errors as well) from the covariates (usually related with the stimuli presented on the experiment) are used as an input product for the group stage. In this work, we present a similar approach, but instead of a GLM, we use a Matrix-Variate Dynamic Linear Model (MVDLM) in order to incorporate the temporal and spatial structures usually present in fMRI data. The description of the modeling at the individual stage is presented in \cite{jimnez2019assessing}. In this work, we mainly focus on the description of the group stage analysis assuming that we already have available the posterior distributions obtained at the individual stage.  

In the next section, we present the statistical procedure to both detections of brain activation patterns from one single group and the comparison of activation strengths between two groups. In section \ref{sec:2}, we present one practical example from an experiment exploring brain activation in the temporal cortex. Finally, in section \ref{sec:4} some concluding remarks are presented.

\section{fMRI Group Analysis}
\label{sec:conc}
Let's suppose we have $N_g$ subjects in $g$ different groups, for which there is a need to explore and/or characterized some aspects related to their brain functioning. The most common cases in the fMRI literature are for $g=1,2$, when either there is an interest for the pattern of activation from a single group or the comparison of activation strength between groups cases and controls. Suppose also that the model (\ref{sec2:equ1}) is fitted at the voxel level for every subject from each group as it is performed in \cite{jimnez2019assessing}.

\begin{equation} \label{sec2:equ1}
\begin{array}{lccl}
\text{Observation:}\ &\mathbf{Y}_{\scaleto{vt\mathstrut}{7pt}}^{\scaleto{(z)\mathstrut}{7pt}} & =& \mathbf{F}^{'}_t\mathbf{\Theta}_{\scaleto{vt\mathstrut}{7pt}}^{\scaleto{(z)\mathstrut}{7pt}} + \bm{\nu}_{\scaleto{vt\mathstrut}{7pt}}^{\scaleto{'(z)\mathstrut}{7pt}} \\
 \text{Evolution:}\ &\mathbf{\Theta}_{\scaleto{vt\mathstrut}{7pt}}^{\scaleto{(z)\mathstrut}{7pt}}& =& \mathbf{G}_t\mathbf{\Theta}_{\scaleto{vt-1\mathstrut}{7pt}}^{\scaleto{(z)\mathstrut}{7pt}} + \mathbf{\Omega}_{\scaleto{vt\mathstrut}{7pt}}^{\scaleto{(z)\mathstrut}{7pt}},\\
\end{array}
\end{equation} 

Where $v=[i,j,k]$ represent the voxel position in the brain image and $z = 1, \ldots, N_g$. Thus, the posterior distribution $p(\mathbf{\Theta}_{\scaleto{vt\mathstrut}{7pt}}^{\scaleto{(z)\mathstrut}{7pt}}|\mathbf{Y}_{\scaleto{vt\mathstrut}{7pt}}^{\scaleto{(z)\mathstrut}{7pt}})$ is given by 

\begin{equation}\label{sec2:equ2}
(\mathbf{\Theta}_{vt}^{(z)}|D_{vt}^{(z)})\sim T_{n_t}[\mathbf{m}_{vt}^{(z)}, \mathbf{C}_{vt}^{(z)}, \mathbf{S}_{tv}^{(z)}].
\end{equation}

In the contex of this modeling the $p\times q$ parameter $\mathbf{\Theta}_{\scaleto{vt\mathstrut}{7pt}}^{\scaleto{(z)\mathstrut}{7pt}}$ brings information about the brain activation for subject $z$, at position $v$ and time $t$. Specifically, when $\{\mathbf{\Theta}_{\scaleto{vt\mathstrut}{7pt}}^{\scaleto{(z)\mathstrut}{7pt}}>0\}$ it is interpreted as a match between the expected (the components of $\mathbf{F}^{'}_t$) and observed BOLD ($\mathbf{Y}_{\scaleto{vt\mathstrut}{7pt}}^{\scaleto{(z)\mathstrut}{7pt}}$) responses, which simply means a brain activation. In that sense, we would like to combine that information among the subjects and compute a new measure which would inform about the activation pattern of the entire group. Thus, in order to do so, we follow a similar approach as in \cite{beckmann2003general}, where $\mathbf{\Theta}_{\scaleto{vt\mathstrut}{7pt}}^{\scaleto{(g)\mathstrut}{7pt}} = \frac{1}{N_g} \sum_{z} \mathbf{\Theta}_{\scaleto{vt\mathstrut}{7pt}}^{\scaleto{(z)\mathstrut}{7pt}}$ represents the average group activation at positon $v$ and time $t$. However, given that the distribution of the original variables is matrix T distribution, dealing with the resulting distribution from the new variable $\mathbf{\Theta}_{\scaleto{vt\mathstrut}{7pt}}^{\scaleto{(g)\mathstrut}{7pt}}$ can be cumbersome. Thus, in order to overcome this problem, we take advantage of this reasonable approximation: the posterior distribution of  $\mathbf{\Theta}_{\scaleto{vt\mathstrut}{7pt}}^{\scaleto{(z)\mathstrut}{7pt}}$ when $n_t\geq 30$ can be represented by

\begin{equation}\label{sec2:equ3}
(\mathbf{\Theta}_{vt}^{(z)}|D_{vt}^{(z)}) \widesim{approx} N_{pq}[\mathbf{m}_{vt}^{(z)}, \mathbf{C}_{vt}^{(z)}, \mathbf{S}_{vt}^{(z)}].
\end{equation}

Then, from the properties of the matrix variate distribution \citep{gupta2018matrix}, the distribution of the average group activation is given by 

\begin{equation}\label{sec2:equ4}
(\mathbf{\Theta}_{vt}^{(g)}|D_{vt}^{(g)}) \widesim{approx} N_{pq} \left[ \frac{1}{N_g}\sum_{z}\mathbf{m}_{vt}^{(z)}, \frac{1}{N_g^2}\sum_{z}\mathbf{C}_{vt}^{(z)}, \frac{1}{N_g^2}\sum_{z}\mathbf{S}_{vt}^{(z)} \right].
\end{equation}

Having the distribution (\ref{sec2:equ4}) available, the group activation at voxel $v$ can be assessed by using any of the algorithms we define in \cite{jimnez2019assessing}. From the posterior distribution (\ref{sec2:equ4}), we define three different probability distributions obtained from the components of $\mathbf{\Theta}_{vt}^{(g)}$:

\begin{equation}\label{sec2:equ5}
\text{Marginal group effect:} \ \ \bar{\theta}_{\scaleto{vt,l\mathstrut}{6pt}}^{*(g)} \sim N(\bar{m}_{\scaleto{vt,l\mathstrut}{6pt}}^{*(g)}, \bar{S}_{\scaleto{vt,l\mathstrut}{6pt}}^{(g)}),
\end{equation}

\begin{equation}\label{sec2:equ6}
\text{Average cluster group effect:} \ \ \bar{\bar{\theta}}_{\scaleto{vt,l\mathstrut}{6pt}}^{(g)}\sim N(\bar{\bar{m}}_{\scaleto{vt,l\mathstrut}{6pt}}^{(g)}, \bar{\bar{S}}_{\scaleto{t,l\mathstrut}{6pt}}^{(g)}),
\end{equation}

\begin{equation}\label{sec2:equ7}
\text{Joint group effect:} \ \ \boldsymbol{\bar{\theta}}_{\scaleto{vt,l\mathstrut}{6pt}}^{(g)}\sim N_q(\boldsymbol{\bar{m}}_{\scaleto{vt,l\mathstrut}{6pt}}^{(g)}, \boldsymbol{\bar{S}}_{\scaleto{t\mathstrut}{6pt}}^{(g)}),
\end{equation}

where $l=1, \ldots, p$. In \cite{jimnez2019assessing}, the interested reader can find a more detailed explanation on how distributions (\ref{sec2:equ5}), (\ref{sec2:equ6}) and (\ref{sec2:equ7}) are obtained.

\subsection*{Algorithms for Group Analysis}

In this section we present the versions for group analysis of the FEST, FSTS and FFBS algorithms. We also present an aditional algorithm, which rely on outputs from those three algorithms at the individual level. 

\subsubsection*{FEST algorithm}

\begin{algorithm}
\caption{Forward Estimated Trajectories Sampler}\label{euclid}
\begin{algorithmic}[1]
\Procedure{ $ \textsf{for}\ \ k=1\ldots N_{simu}$, $p(\theta^{(g)}_{vt,l}|D_{vt}^{(g)})$ being any of (\ref{sec2:equ5}), (\ref{sec2:equ6}) or (\ref{sec2:equ7})}{}
\State Draw $\theta^{(g,k)}_{vt,l}$ from $p(\theta^{(g)}_{vt,l}|D_{vt}^{(g)})$ for $t=1,\ldots,T$ and $l=1, \ldots, p$
\State Draw $\nu_{vt}^{(g,k)}$ from $N_q[0,\boldsymbol{s}_{\scaleto{vt\mathstrut}{6pt}}^{(g)}]$ for $t=1,\ldots,T$
\State Compute $\tilde{Y}^{(g,k)}_{vt}= \sum \limits_{l=1}^{p}\theta^{(g,k)}_{vt,l}x^{(g)}_l(t) +\nu_{vt}^{(g,k)}$ for $t=1,\ldots,T$
\State Compute $p(\theta^{(g,k)}_{vt,l}|\tilde{D}_{vt}^{(g)})$ for $t=1,\ldots,T$ and $l=1, \ldots, p$, where $\tilde{D}_{vt}^{(g)} = \{\tilde{Y}^{(g,k)}_{v1}, \ldots, \tilde{Y}^{(g,k)}_{vt}\}$
\State Let $\gamma_{v,l}^{(g,k)}=\{E(\theta^{(g,k)}_{v1,l}|\tilde{D}_{v1}^{(g)}), \ldots, E(\theta^{(g,k)}_{vT,l}|\tilde{D}_{vT}^{(g)})\}$ for $l=1, \ldots, p$
\EndProcedure
\end{algorithmic}
\end{algorithm}

Thus, a measure of evidence for group activation related to stimulus $l$ at position $v$ is given by
\begin{equation}\label{moteCarlo1}
p(\gamma_{v,l}^{(g)}>0) = E(\mathds{1}_{(\gamma_{v,l}^{(g)}>0)})\approx \frac{\sum\limits_{k=1}^{N_{simu}} \mathds{1}_{(\gamma_{v,l}^{(g,k)}>0)}}{N_{simu}}.
\end{equation}

One limitation of this algorithm is that it depends on $x^{(g)}_l(t)$ being the same for all the subjects. Despite most of the fMRI group experiment meeting this requirement, there are some designs where random sequences of stimuli are presented, so in cases like these, the form of the expected BOLD response ($x^{(g)}_l(t)$) will be different among the subjects. The FSTS and FFBS algorithms can help to overcome this specific limitation since they do not depend on $x^{(g)}_l(t)$.

\subsubsection*{FSTS algorithm}

\begin{algorithm}
\caption{Forward State Trajectories Sampler}\label{euclid2}
\begin{algorithmic}[1]
\Procedure{$ \textsf{for}\ \ k=1\ldots N_{simu}$}{}
\State Compute $\mathbf{W}_{vt}^{(g)} = \mathbf{B_t}\mathbf{C}_{v,t-1}^{(g)}\mathbf{B_t} - \mathbf{C}_{v,t-1}^{(g)}$ for $t=1,\ldots,T$
\State Draw $\mathbf{\Omega}_{vt}^{(g,k)}$ from $ N_{pq}[\mathbf{0}, \mathbf{W}_{vt}^{(g)}, \boldsymbol{S}_{\scaleto{vt\mathstrut}{6pt}}^{(g)}]$ for $t=1,\ldots,T$
\State Draw $\mathbf{\Theta}_{v,t-1}^{(g,k)}$ from $N_{pq}[\mathbf{m}_{v,t-1}^{(g)}, \mathbf{C}_{v,t-1}^{(g)}, \mathbf{S}_{v,t-1}^{(g)}]$ for $t=1,\ldots,T$
\State Compute $\mathbf{\Theta}_{vt}^{(g,k)}=\mathbf{\Theta}_{v,t-1}^{(g,k)}+\mathbf{\Omega}_{vt}^{(g,k)}$ for $t=1,\ldots,T$
\State Let $\hat{\mathbf{\Theta}}^{(g,k)}_{v} = \{ \mathbf{\Theta}_{v1}^{(g,k)}, \ldots, \mathbf{\Theta}_{vT}^{(g,k)}\}$

\EndProcedure
\end{algorithmic}
\end{algorithm}

Using Monte Carlo integration as in (\ref{moteCarlo1}), one can test brain activation just by taking the appropriate components from $\hat{\mathbf{\Theta}}^{(g,k)}_{v}$ to compute whichever marginal, average or joint effects.

\subsubsection*{FFBS algorithm}

The FFBS algorithm at the individual level rely on the filtered distribution $p(\mathbf{\Theta}_{v,t-j}^{(z)}|\mathbf{\Theta}_{v,t-j+1}^{(z)},\mathbf{\Sigma}^{(z)},D_{vt}^{(z)})$, which under the assumptions and prior distributions set for model (\ref{sec2:equ1}) is given by

\begin{equation}\label{sec2:equ8}
 N_{pq}(\mathbf{m}_{vj}^{*(z)}, \mathbf{C}_{vj}^{*(z)}, \mathbf{\Sigma}^{(z)}).
\end{equation}

In order to define the FFBS algorithm for the group case, we just adapt the distribution (\ref{sec2:equ8}) by replacing its parameters $\mathbf{m}_{vj}^{*(z)}$, $\mathbf{C}_{vj}^{*(z)}$ and $\mathbf{\Sigma}^{(z)}$ for 
$\mathbf{m}_{vj}^{*(g)}$, $\mathbf{C}_{vj}^{*(g)}$ and $\mathbf{\Sigma}^{(g)}$ respectively. In this way, the adapted filtered distribution for the group case is given by 

\begin{equation*}
(\mathbf{\Theta}_{v,t-j}^{(g)}|\mathbf{\Theta}_{v,t-j+1}^{(g)},\mathbf{\Sigma}^{(g)},D_{vt}^{(g)}) \sim N_{pq}(\mathbf{m}_{vj}^{*(g)}, \mathbf{C}_{vj}^{*(g)}, \mathbf{\Sigma}^{(g)}),
\end{equation*}

where $\mathbf{m}_{vj}^{*(g)} = \mathbf{m}_{v,t-j}^{(g)} + \mathbf{C}_{v,t-j}^{(g)}(\mathbf{B_t}\mathbf{C}_{v,t-j}^{(g)}\mathbf{B_t})^{-1}(\mathbf{\Theta}_{v,t-j+1}^{(g)} - \mathbf{m}_{v,t-j}^{(g)})$ and $\mathbf{C}_{vj}^{*(g)}=\mathbf{C}_{v,t-j}^{(g)} - \mathbf{C}_{v,t-j}^{(g)}(\mathbf{B_t}\mathbf{C}_{v,t-j}^{(g)}\mathbf{B_t})^{-1}\mathbf{C}_{v,t-j}^{(g)}$.

\begin{algorithm}
\caption{Forward-filtering-backward-sampling}\label{euclid3}
\begin{algorithmic}[1]
\Procedure{$ \textsf{for}\ \ k=1\ldots N_{simu}$}{}
\State Draw $\mathbf{\Sigma}^{(g,k)}$ from $W^{-1}_{n_t}(\mathbf{S}_{t}^{(g)})$ 
\State For $j=0$ draw $\mathbf{\Theta}_{v,t}^{(g,k)}$ from $N_{pq}(\mathbf{m}_{vj}^{(g)}, \mathbf{C}_{vj}^{(g)}, \mathbf{\Sigma}^{(g,k)})$
\State For $1\leq j<t$ draw $\mathbf{\Theta}_{v,t-j}^{(g,k)}$ from $p(\mathbf{\Theta}_{v,t-j}^{(g)}|\mathbf{\Theta}_{v,t-j+1}^{(g)},\mathbf{\Sigma}^{(g)},D_{vt}^{(g)})$
\State Let $\hat{\mathbf{\Theta}}^{(g,k)}_{v} = \{ \mathbf{\Theta}_{v1}^{(g,k)}, \ldots, \mathbf{\Theta}_{vt}^{(g,k)}\}$
\EndProcedure
\end{algorithmic}
\end{algorithm}

From the simulated on-line trajectories $\hat{\mathbf{\Theta}}^{(g,k)}_{v}$ one can compute an activation evidence as in the case of the FSTS algorithm.

\subsubsection*{AG-algorithm}
Average Group algorithm (AG-algorithm) is just another way to simulate on-line trajectories of the state parameter, which are used to assess voxel activation at the group level. However, instead of using the distribution of the average group effect (\ref{sec2:equ4}), the AG-algorithm samples on-line estimated trajectories at the individual level and then compute an average group on-line trajectory at voxel $v$. What is intended with this approach is to allow more variability among subjects to be taken into account when computing group activation effects.

\begin{algorithm}[H]
\caption{AG-algorithm: $ \textsf{for}\ \ k=1\ldots N_{simu}$}\label{euclid4}
\begin{algorithmic}[1]
\Procedure{}{}
\State Sample one on-line trajectory using any algorithm from \{FEST, FFBS, FSTS\} for $z=1,\ldots, N_g$ at voxel $v$.
\State Make $
\beta_{v}^{(z,k)}=
\begin{cases}
\gamma_{v,l}^{(z,k)},\ \ \textbf{if FEST is the chosen sampler}\\
\hat{\mathbf{\Theta}}_{v}^{(z,k)}, \ \ \textbf{if either FSTS or FFBS is the chosen sampler}
\end{cases}
$
\State Compute $\beta_{v}^{(g,k)}=\frac{1}{N_g}\sum \limits_{z=1}^{_Ng}\beta_{v}^{(z,k)}$
\EndProcedure
\end{algorithmic}
\end{algorithm}

Therefore, a measure of evidence for group activation at voxel $v$ is given by:

\begin{equation}\label{moteCarlo1}
p(\beta_{v}^{(g)}>0) = E(\mathds{1}_{(\beta_{v}^{(g)}>0)})\approx \frac{\sum\limits_{k=1}^{N_{simu}} \mathds{1}_{(\beta_{v}^{(g,k)}>0)}}{N_{simu}}.
\end{equation}

\section{Example}
\label{sec:2}

\begin{figure}[H]
\centering
\includegraphics[width=.80\textwidth]{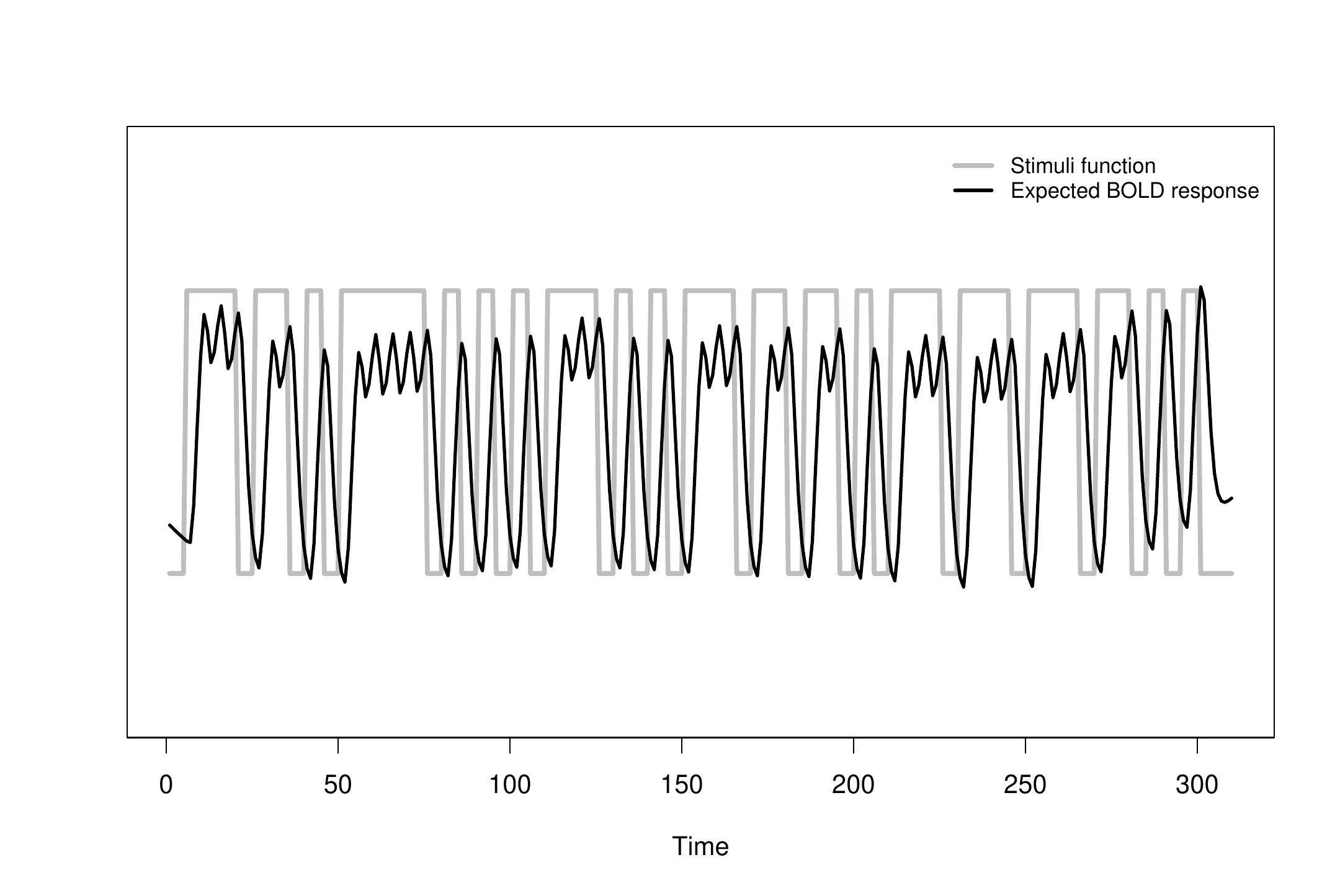}
\caption{\label{fig:1} Merged stimuli function for voice and non-voice sounds with its respective expected BOLD response function for the "voice localizer" example.}
\end{figure}

In order to present a practical example using the method proposed in this work, we use the data from \cite{pernet2015human}, which is openly available on \textbf{OpenNeuro} \citep{gorgolewski2017openneuro}. In that work, they perform an fMRI experiment where a group of $n=207$ individuals is submitted to sound stimulation composed by human voices and non-human sounds in order to identify or characterize the brain regions involved in the recognition of human voice sounds. For the sake of simplicity, we merge both human and non-human sounds in only one block design (figure \ref{fig:1}) and take just 21 (sub-001:sub-021) from the 207 participants in the study. Our main aim with this example is to evaluate the capacity of our method to identify brain activation due to sound stimulation. It is also worth mentioning that brain images from all the subjects are converted to the MNI atlas \citep{brett2002problem} to make their brains comparable, which is a standard procedure in fMRI group analysis.

\begin{table}[H]
\begin{figure}[H]
  \centering
\begin{center}
\begin{tabular}{cc}
Marginal-FEST&Joint-FEST\\
\includegraphics[width=.35\textwidth]{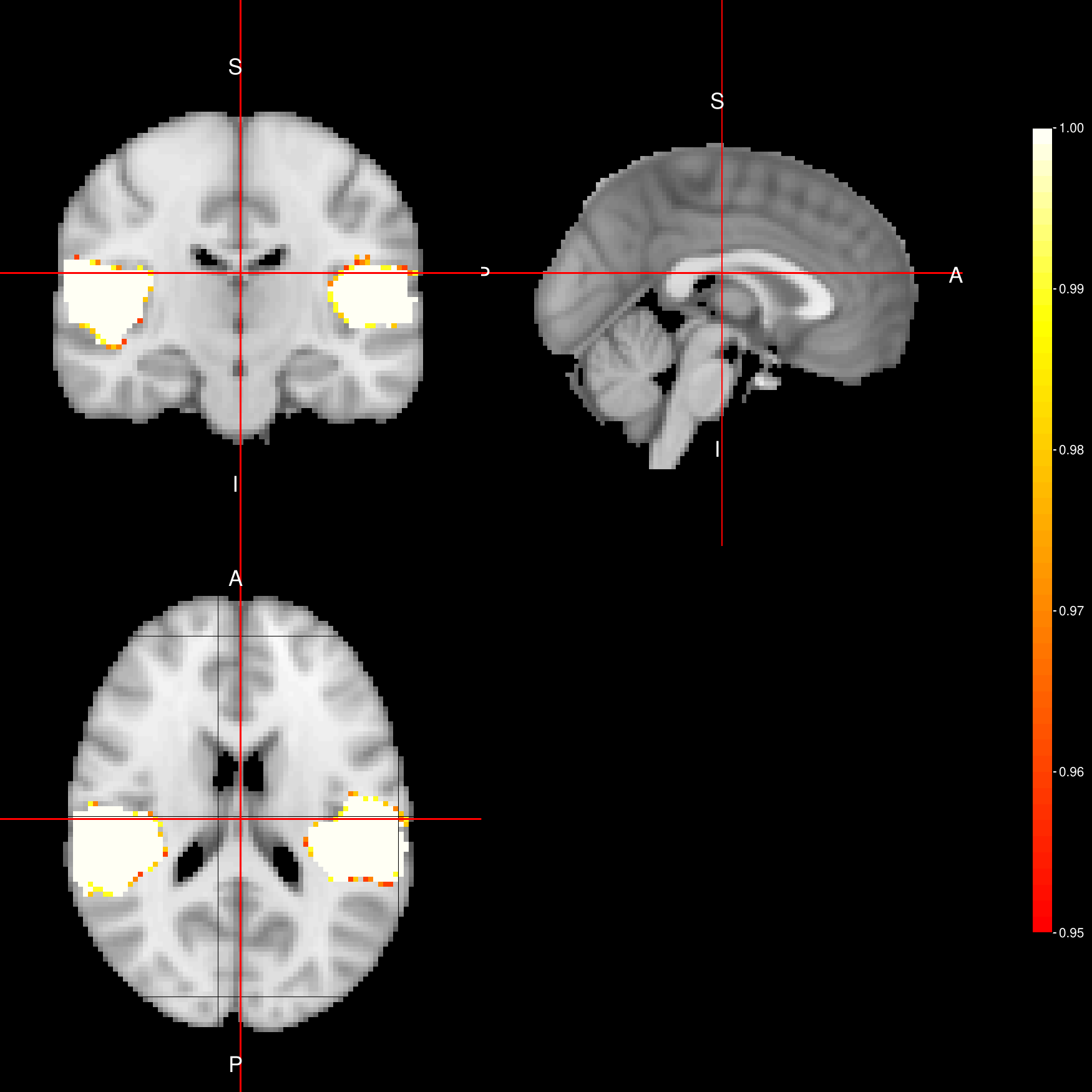}&\includegraphics[width=.35\textwidth]{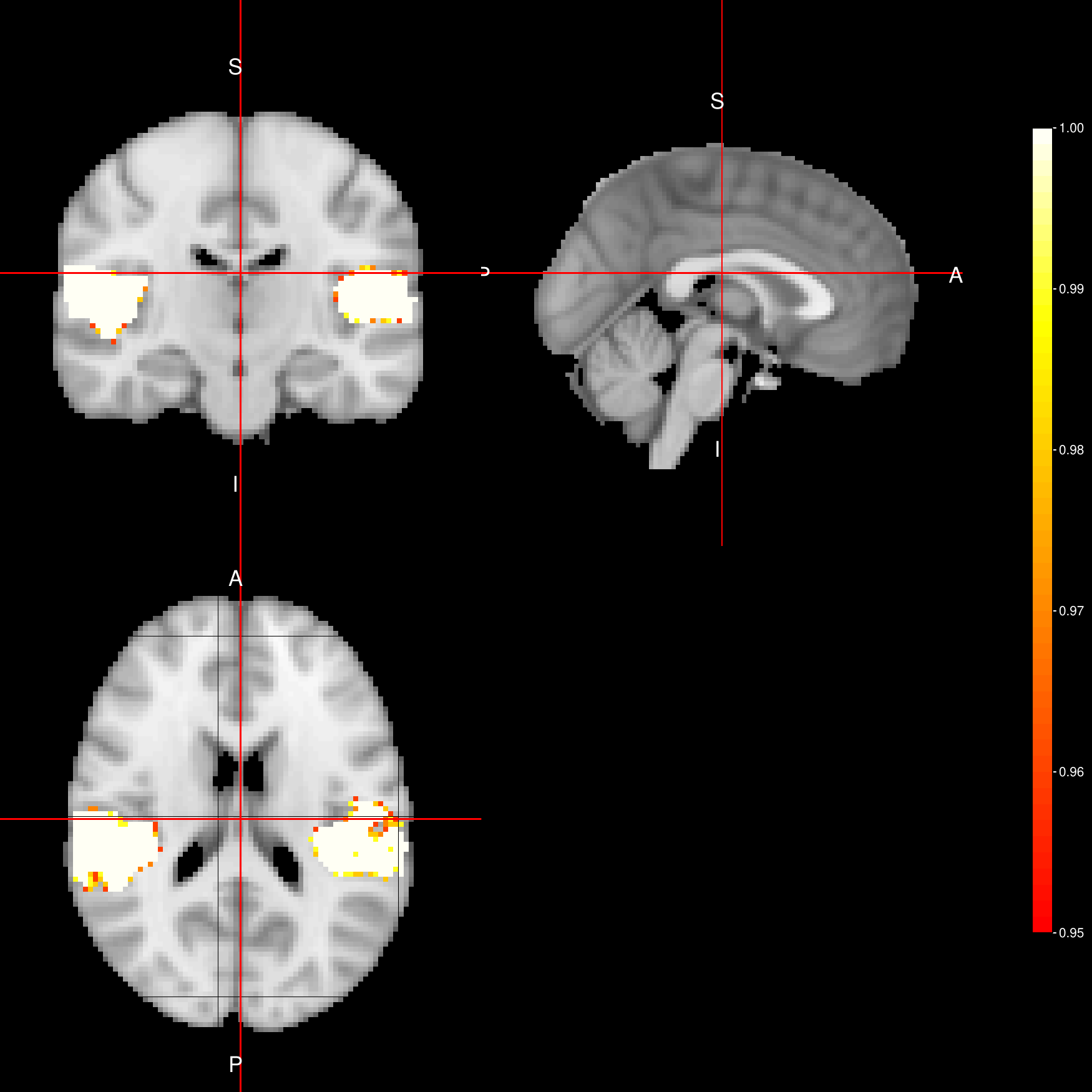}\\
\multicolumn{2}{c}{ACE-FEST}\\
\multicolumn{2}{c}{\includegraphics[width=.35\textwidth]{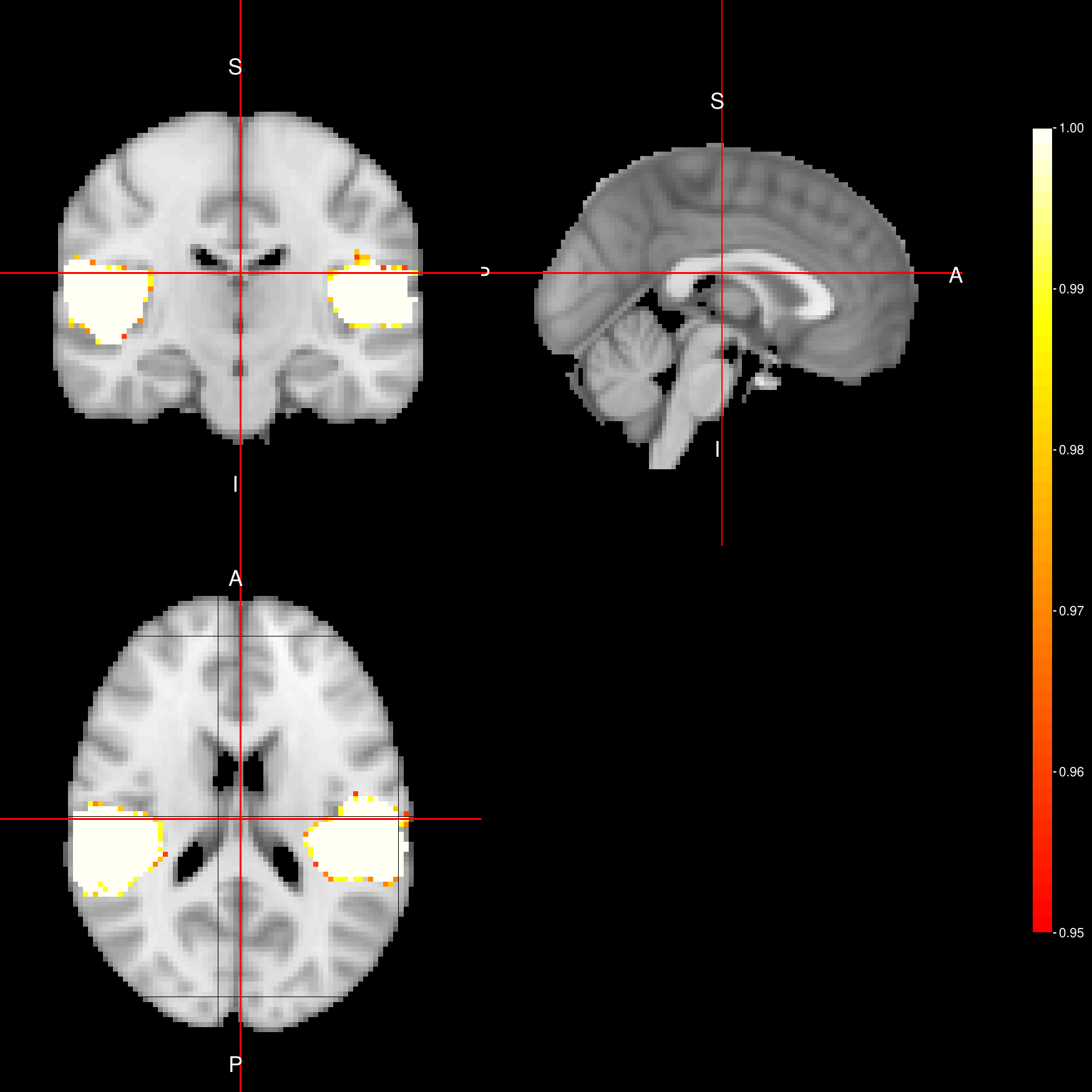}}
\end{tabular}
\end{center}
  \caption{Activation Maps for the "voice localizer" example obtained when using the FEST algorithm under three different distributions (Marginal, Joint and LTT) related to the state parameter.}
  \label{fig2} 
\end{figure}
\end{table}

\begin{table}[H]
\begin{figure}[H]
  \centering
\begin{center}
\begin{tabular}{cc}
Marginal-FFBS&Joint-FFBS\\
\includegraphics[width=.35\textwidth]{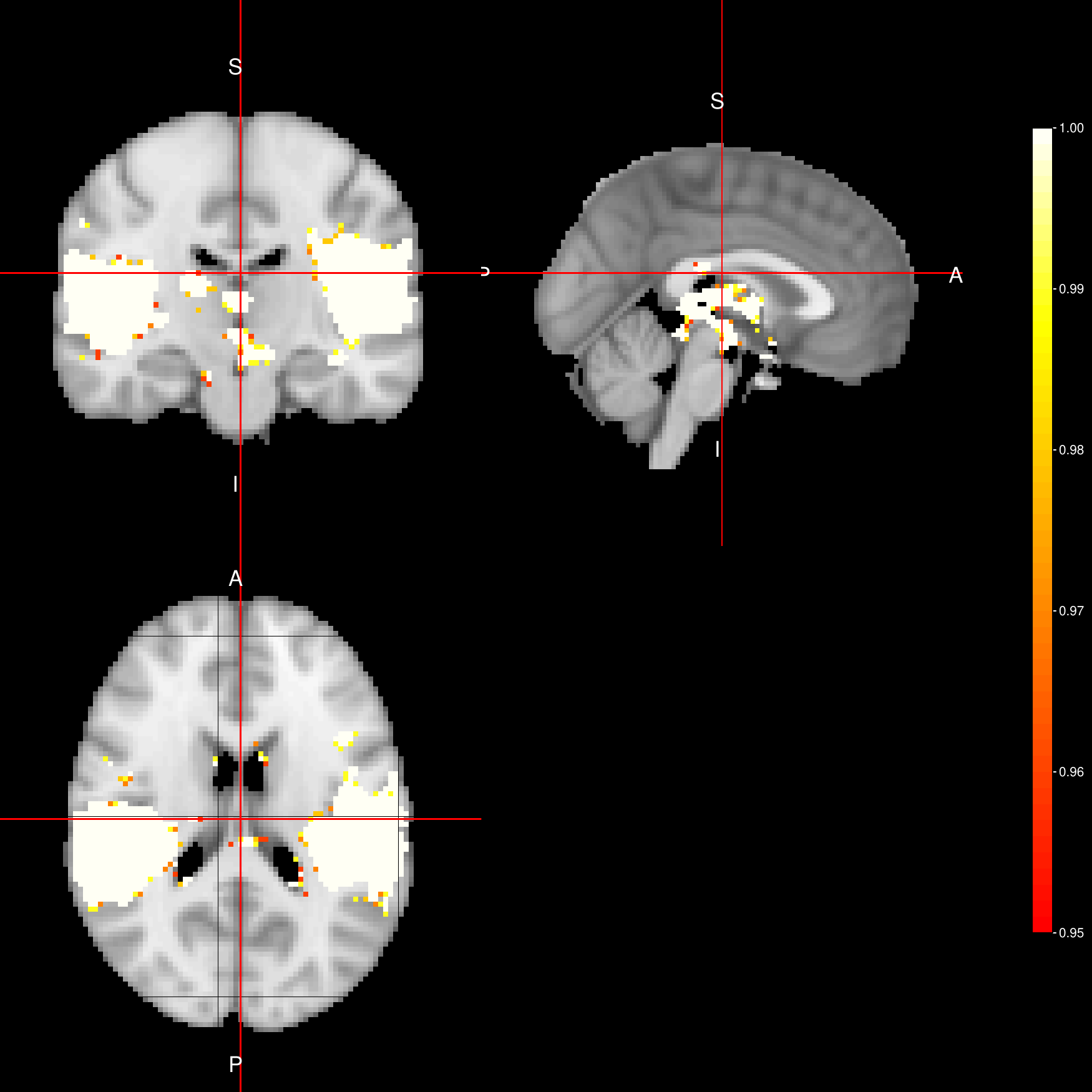}&\includegraphics[width=.35\textwidth]{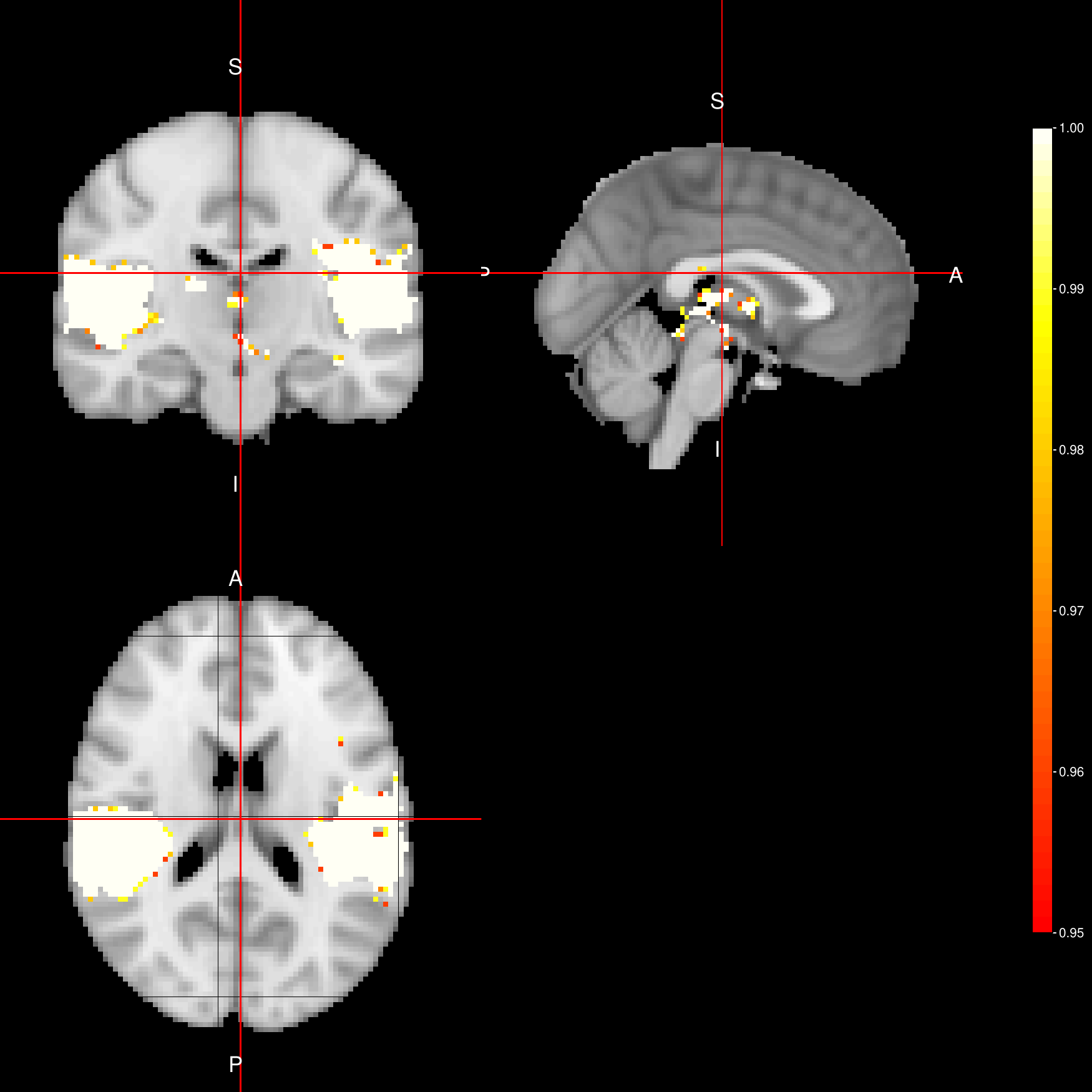}\\
\multicolumn{2}{c}{ACE-FFBS}\\
\multicolumn{2}{c}{\includegraphics[width=.35\textwidth]{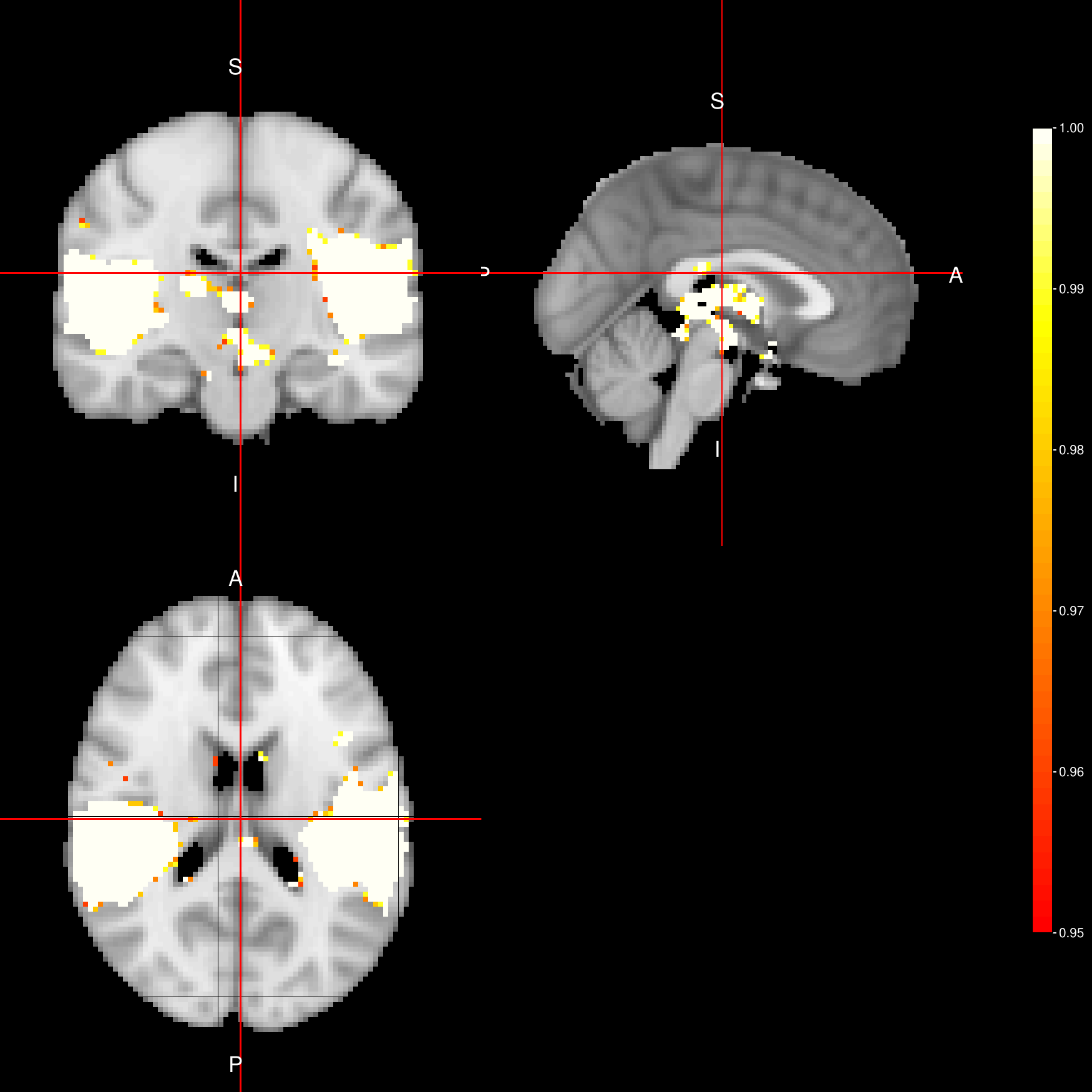}}
\end{tabular}
\end{center}
  \caption{Activation Maps for the "voice localizer" example obtained when using the FFBS algorithm under three different distributions (Marginal, Joint and LTT) related to the state parameter.}
  \label{fig3} 
\end{figure}
\end{table} 

From figures \ref{fig2}, \ref{fig3} and \ref{fig4}, we can see the activation maps obtained for the "voice localizer" experiment using the method proposed in this work.  From those images, we can say that the three algorithms (FEST, FFBS and FSTS) under the three different distributions (Marginal, joint and LTT or average distribution) successfully identify the temporal activation due to voice and non-voice sounds stimulation, nevertheless there are some slight differences among those maps worth mentioning. For instance, the maps obtained when using the FFBS algortihm allows for the identification of a broader activated region from the temporal cortex, however, on the other hand, it allows activations to appear (false-positive activations) on brain regions that should not be involved with this "voice localizer" experiment. On the other hand, more conservative results seem to be obtained when using FEST and FSTS algorithms, but with less false activations.

\begin{table}[H]
\begin{figure}[H]
  \centering
\begin{center}
\begin{tabular}{cc}
Marginal-FSTS&Joint-FSTS\\
\includegraphics[width=.35\textwidth]{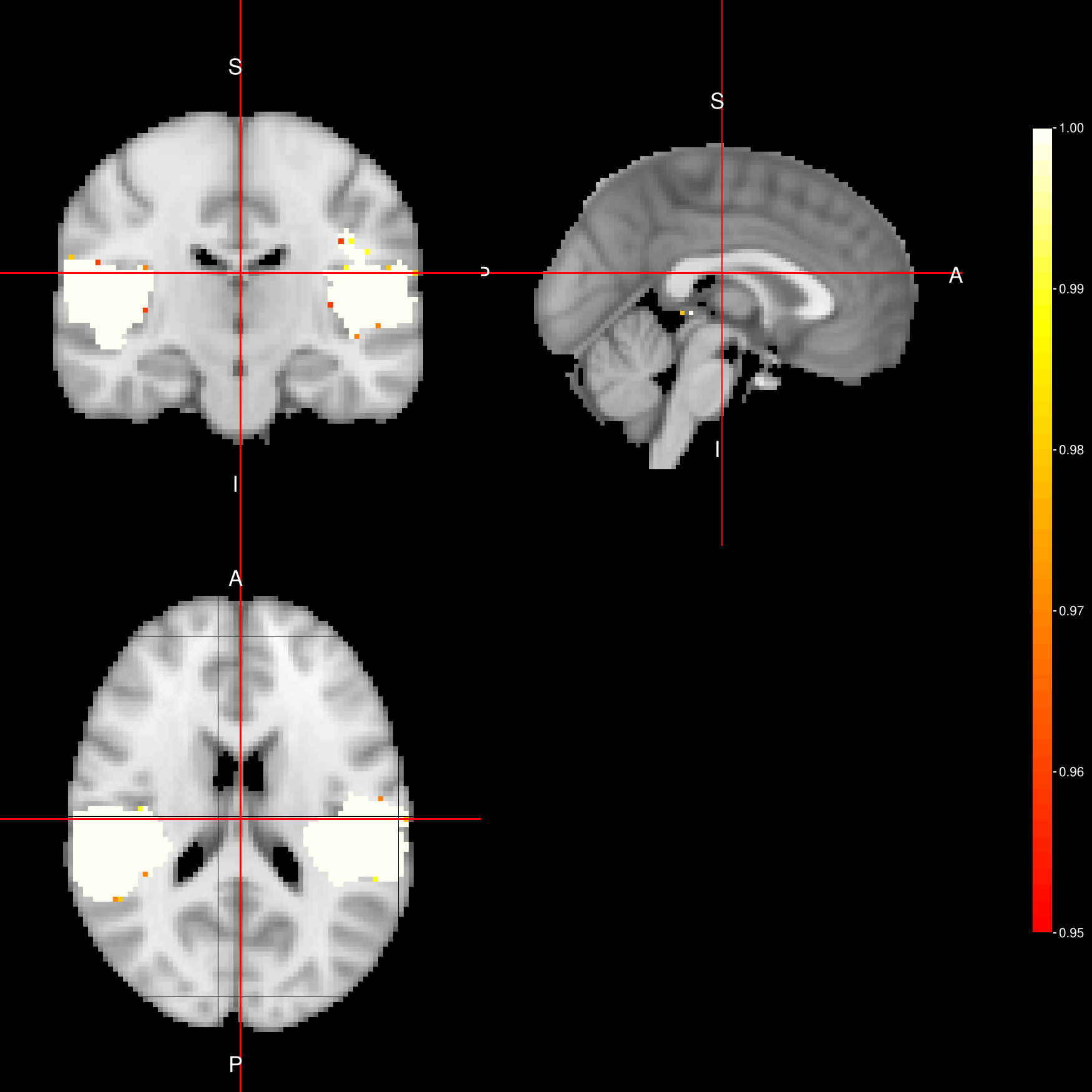}&\includegraphics[width=.35\textwidth]{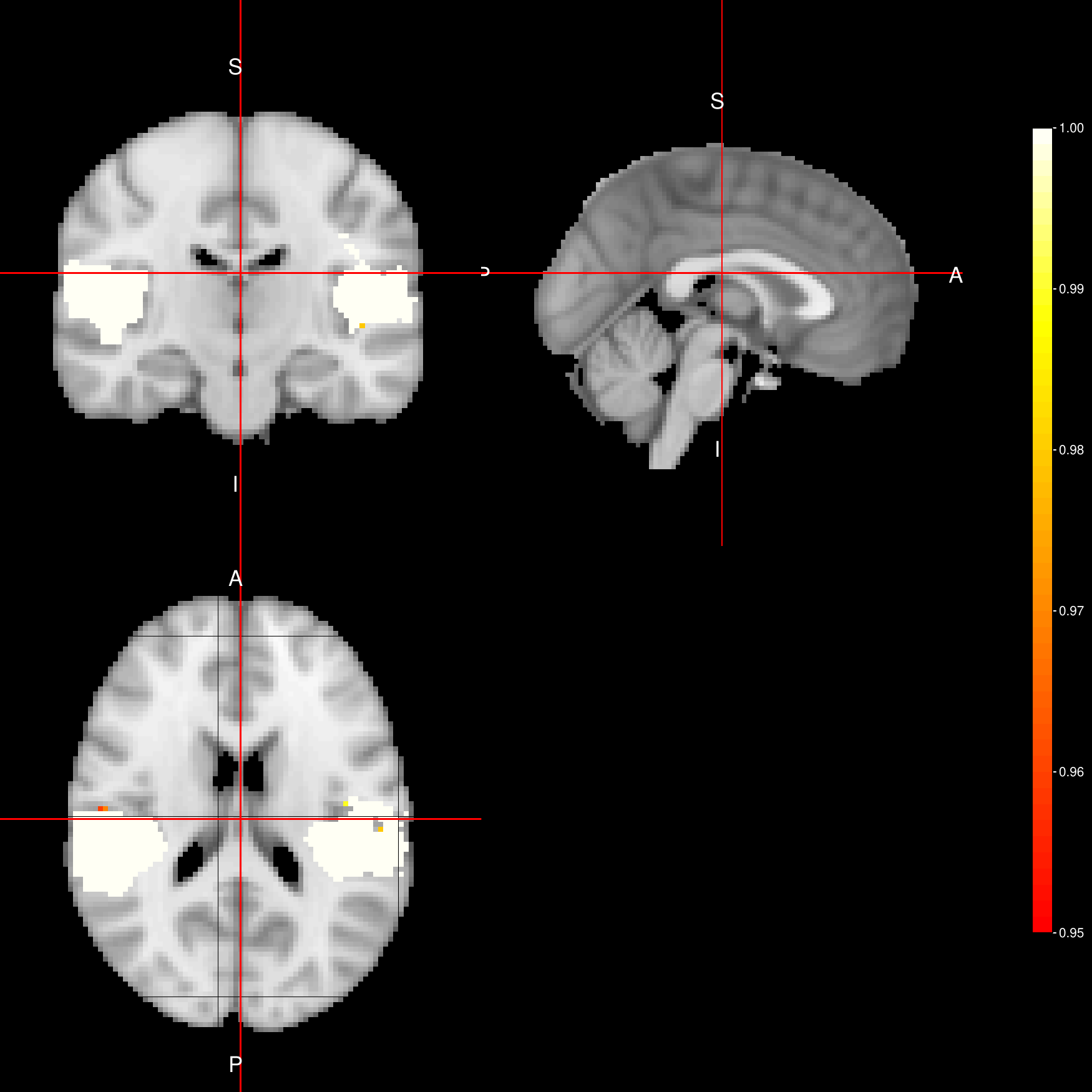}\\
\multicolumn{2}{c}{ACE-FSTS}\\
\multicolumn{2}{c}{\includegraphics[width=.35\textwidth]{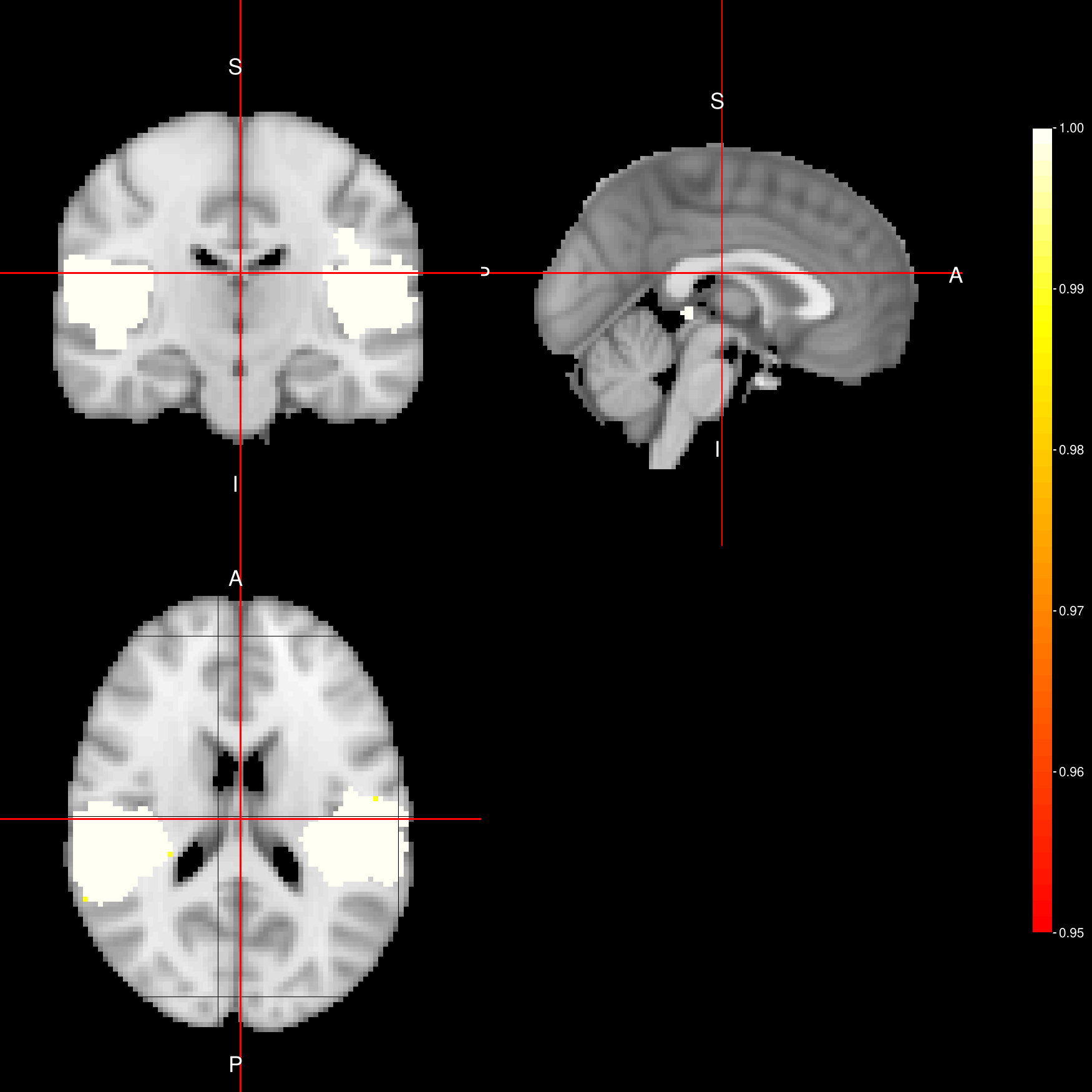}}
\end{tabular}
\end{center}
  \caption{Activation Maps obtained for the "voice localizer" example when using the FSTS algorithm under three different distributions (Marginal, Joint and LTT) related to the state parameter.}
  \label{fig4} 
\end{figure}
\end{table}

\begin{table}[H]
\begin{figure}[H]
  \centering
\begin{center}
\begin{tabular}{cc}
AG-algorithm(ACE-FEST)&AG-algorithm(ACE-FFBS)\\
\includegraphics[width=.35\textwidth]{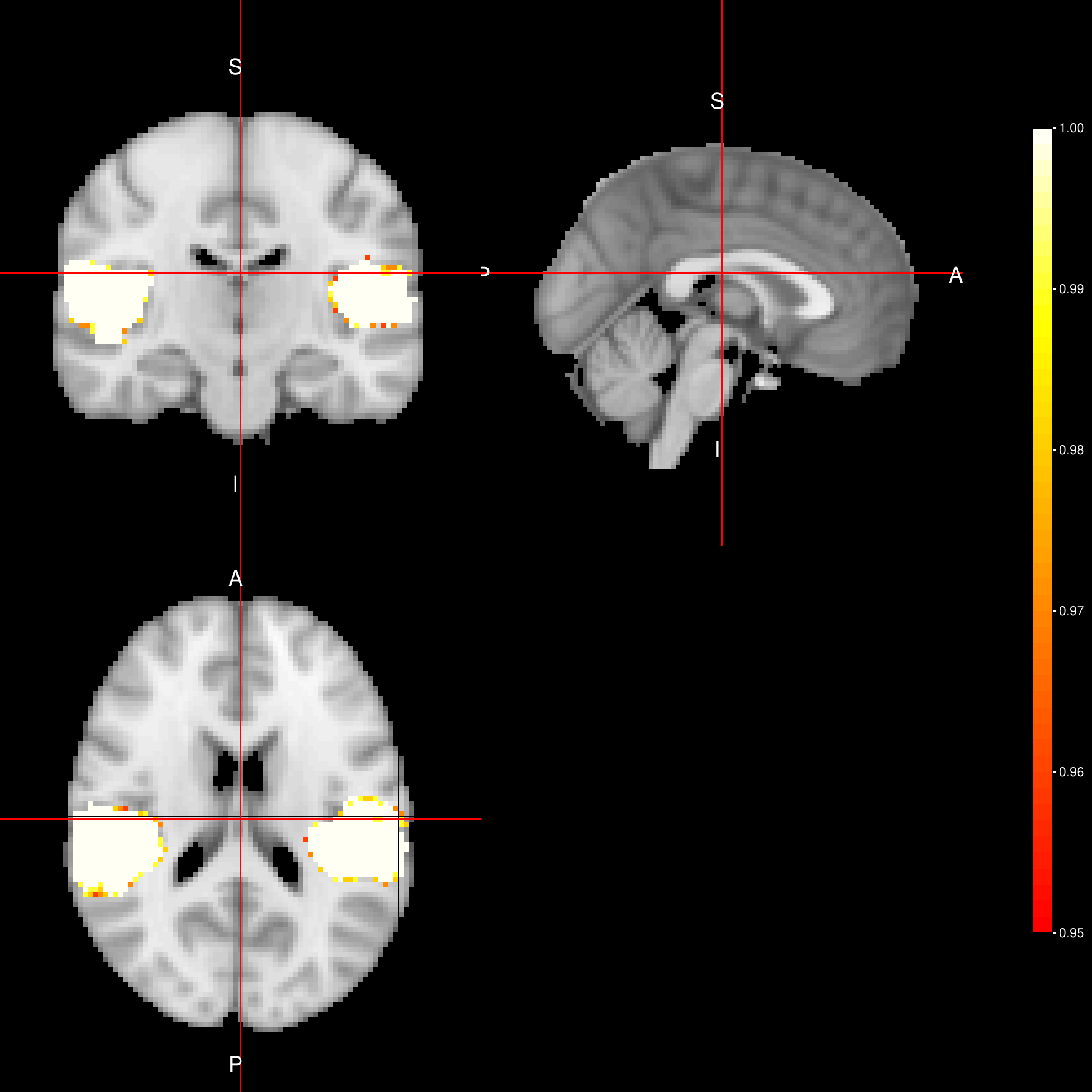}&\includegraphics[width=.35\textwidth]{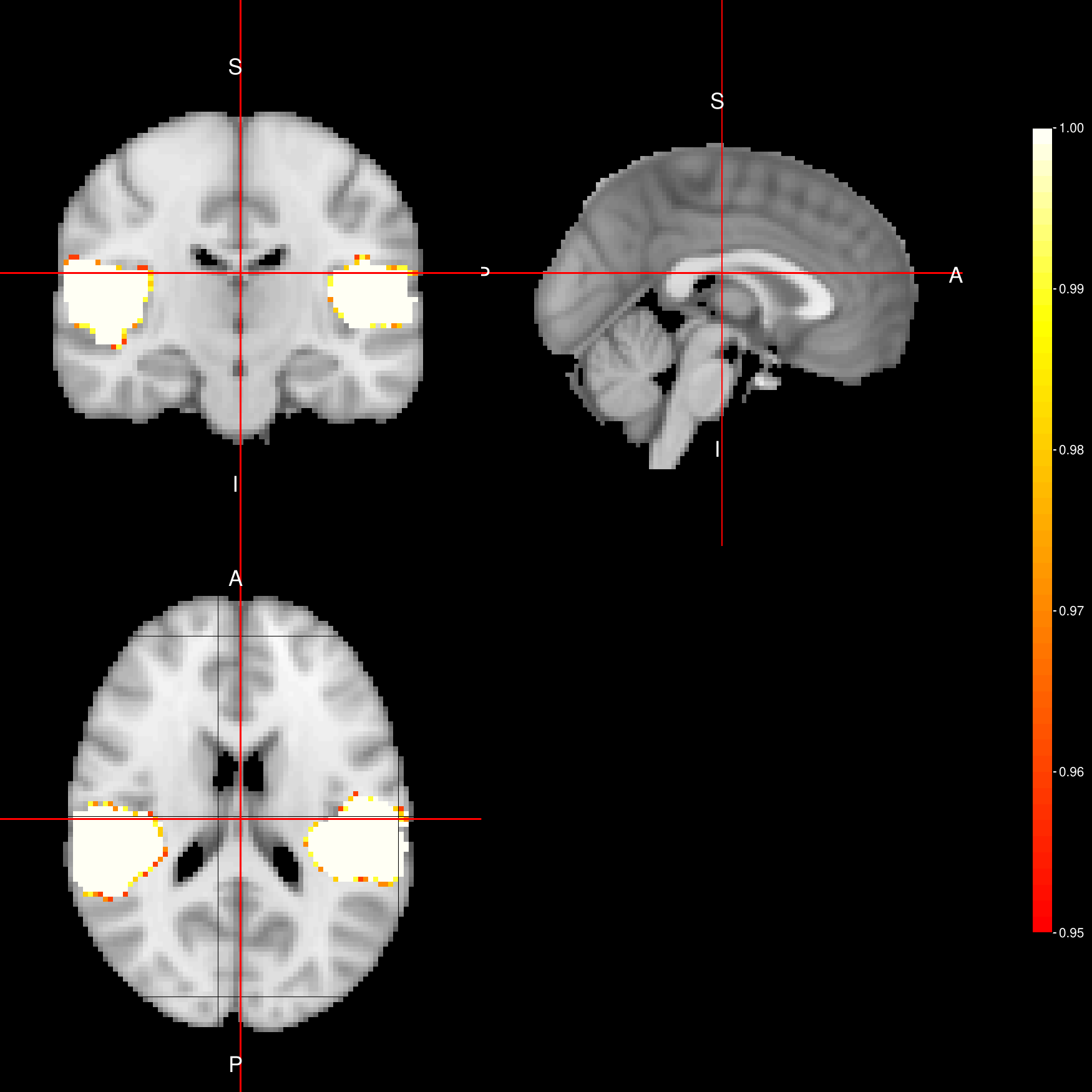}\\
\multicolumn{2}{c}{AG-algorithm(ACE-FSTS)}\\
\multicolumn{2}{c}{\includegraphics[width=.35\textwidth]{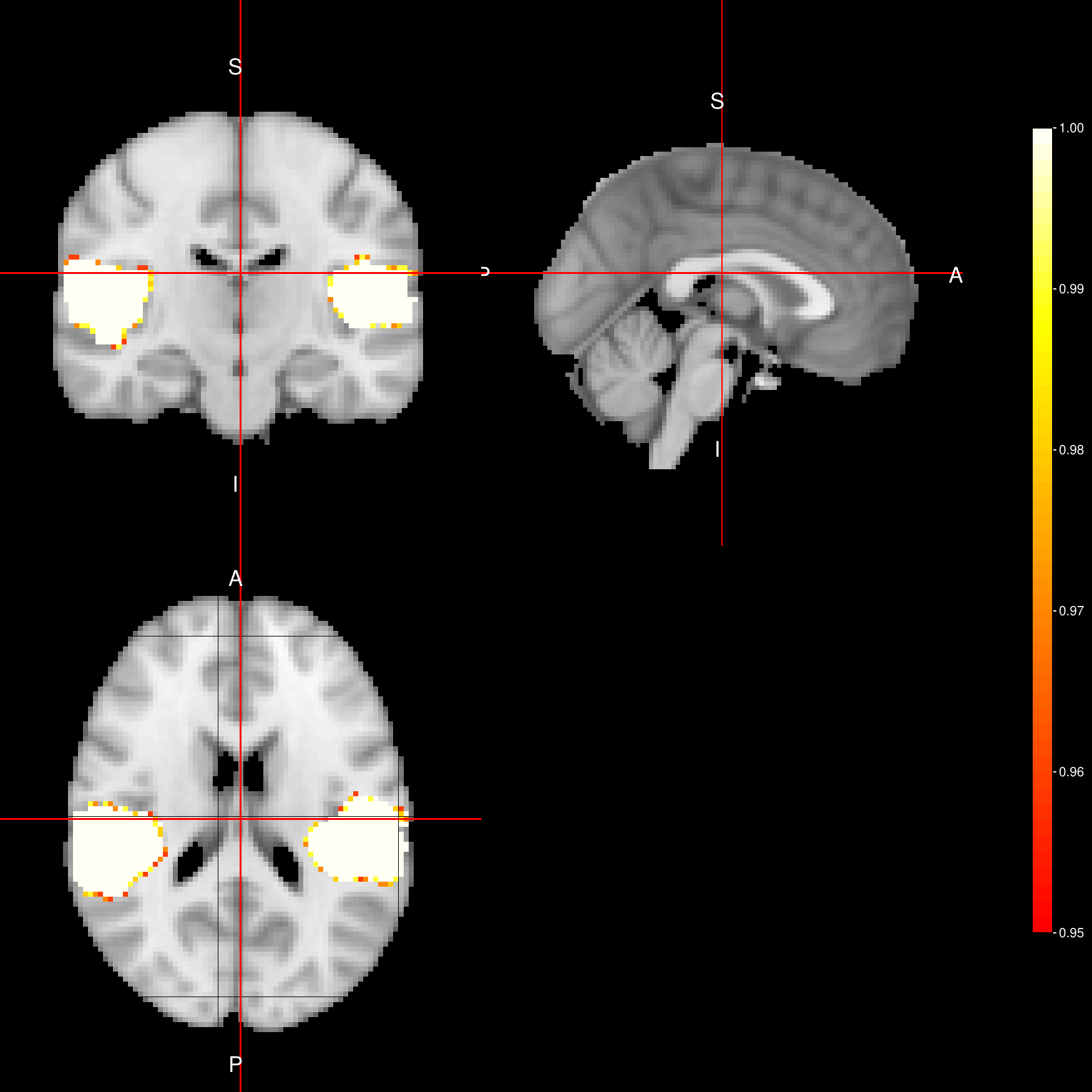}}
\end{tabular}
\end{center}
  \caption{Activation Maps obtained for the "voice localizer" example when using the AG-algorithm}
  \label{fig5} 
\end{figure}
\end{table} 

In figure \ref{fig5}, we can see group activation maps obtained when using the AG-algorithm for every sampler option (FEST, FFBS, FSTS) at the individual level. From this example, we can conclude that with the AG-algorithm it is also possible to identify brain activation when analysing fMRI data for group activation.


\section{Concluding remarks}
\label{sec:4}

In this work, we present a method for fMRI group analysis, which is just a continuation of the method proposed by \cite{jimnez2019assessing} for fMRI data analysis using MDLM at the individual level. It has shown to be very effective when analyzing fMRI data for a group of 21 subjects from a "voice localizer" experiment. We also introduce a new algorithm (AG-algorithm), which allows us to sample on-line tractories of the state parameter from each subject individually, instead of combining the group information into a single posterior distribution (\ref{sec2:equ4}) to sample from. What is intended with this approach is to allow more uncertainty from the intra-subject variability to be taken into account when assessing group voxel activation. Even though, no group comparison is performed in this paper, it can be easily implemented and we just let it as a future work. We also want to stress that other types of analysis, such as group analysis for repeated measures can be easily addressed under this setting thanks to the flexibility of the MDLM. Despite this method being successfully tested with many other fMRI data sets from different types of experiments, a more indepth assessment (as it is made in \cite{jimnez2019assessing}) using real and simulated data must be performed in order to offer a more reliable validation of it.


\begin{description}

\item

\end{description}

\bibliographystyle{unsrtnat}
\bibliography{biblio}

\end{document}